# High-altitude ballooning programme at the Indian Institute of Astrophysics

*Akshata Nayak\*, A. G. Sreejith, M. Safonova and Jayant Murthy*

*Recent advances in balloons and in electronics have made possible scientific payloads at costs accessible to university departments. The primary purpose of high-altitude ballooning programme is to test low-cost ultraviolet payloads for eventual space flight, but to also explore phenomena occurring in the upper atmosphere, including sprites and meteorite impacts, using balloon-borne payloads. This article discusses the results of three tethered balloon experiments carried out at the Centre for Research and Education in Science and Technology (CREST) campus of IIA, Hosakote and our plans for the future. We also describe the stages of payload development for these experiments.*

**Keywords:** Balloons, scientific payloads, tethered flights, upper atmosphere.

NEAR space is the region of the Earth's atmosphere that lies between 20 and 100 km amsl, encompassing the stratosphere, mesosphere and the lower thermosphere. There is now a surge of interest in flying balloons to the edge of near space, largely for recreational purposes, and the internet is filled with videos showing the blackness of space and curvature of the surface of the Earth[1]. We propose to extend this paradigm to fly scientific payloads for a fraction of the cost of a space experiment[2]. Despite the fact that large orbital observatories provide accurate observations and statistical studies of remote and/or faint space sources, small telescopes on-board balloons or rockets are still attractive because they are much cheaper and could still yield substantial scientific output; for example, the first ultraviolet (UV) spectrum of a quasar was obtained during a rocket flight[3]. Balloons are used in many astronomical fields, including cosmic-ray physics, IR astronomy, high-energy and upper-atmospheric physics, and even space biology. Engineering experiments on new space technologies are also conducted using balloons. In 2001, balloons accounted for 9% of all NASA discoveries, according to the 2001 Science News Metrics for NASA[4]. In India, balloons are used by TIFR and PRL to conduct IR and high-energy observations[5,6]. We have begun a new programme with lightweight and low-cost balloons to observe in the near-UV window from 200 to 400 nm that has been yet largely unexplored by the balloon community. This window includes the spectroscopic lines from several key players in atmospheric chemistry such as $SO_2$, $O_3$, BrO and HCHO. It also allows us to observe spectroscopic lines in astronomical sources, for example, in comets and meteors[7]. In addition in India, the high lightning activity during the monsoon months provides opportunities to study UV flashes associated with the transient light events in the upper atmosphere. Such flashes are initiated by electrical discharges between the clouds and ionosphere (10–100 km) and produce the fluorescence radiation of the atmosphere in the 300–400 nm window. They are believed to be related to lightning discharges[8] and it would be important to check this connection by initiating the new area of research using unique conditions provided by the Indian weather.

Commercially available high-altitude balloons provide access to near space at a surprisingly low cost, with the capability to launch payloads of 1–3 kg to altitudes of 35–80 km from the Earth's surface. A flight may last between 3 and 5 h and is terminated when the balloon bursts or the connection between the payload and the balloon is cut. Depending on the weight and type of the balloon, the applications differ. For our purpose, we have selected white latex balloons filled with helium. Two types of latex balloons have been procured for this project from two manufacturers: 1.2 kg lift capacity balloons from Pawan Balloons (Pune, India) and 3 kg lift capacity balloons from Ningbo Yunhuan Electronics Group Co. Ltd (China). While hydrogen gas is approximately 7% more buoyant, we have chosen helium as it has the advantage of being non-flammable. Two helium tanks of 10 $m^3$ capacity have been procured from Sri Vinayaka Gas Agencies, Bangalore. The volume of helium required for a flight depends on its lift capacity. It is known[9,10] that 1 $ft^3$ of helium lifts 28 g; after conversion to a metric system, we find that 2.83 $m^3$ of helium lifts 2800 g, or lifting capacity of helium is 1 $kg/m^3$. For the 1.2 kg balloon

Akshata Nayak is at the Indian Institute of Astrophysics, Bangalore 560 034 and also at the Jain University, Bangalore 562 112, India; A. G. Sreejith, M. Safonova and Jayant Murthy are in the Indian Institute of Astrophysics, Bangalore 560 034, India.
\*For correspondence. (e-mail: akshata@iiap.res.in)

flight with a payload, we calculate the required volume of helium as follows:

| | |
|---|---|
| Balloon weight | = 1200 g |
| Payload weight | = 1000 g |
| Free lift | = 1200 g |
| Total lift | = 3400 g |

The total lift required is 3.4 kg and the required volume of helium must be 3.4 m$^3$. We estimate that we can have two complete flights with one 10 m$^3$ tank of helium. For 3 kg baloon with 1 kg payload, the volume of helium required is 7 m$^3$.

The primary purpose of this article is to show how low-cost balloons can play a role in the building of a scientific experiment and the processes involved therein. After applying for the approvals necessary for a free-flying balloon, we have used the waiting period to develop the payload. We describe the payload development and initial results from three tethered flights in the next two sections, then our plans for the future followed by a summary.

## Development of a payload

Our ultimate goal is to fly a spectrograph to measure atmospheric and astronomical emissions in the near-UV. This requires a stable platform where the data can be stored for later retrieval or transmission. The challenge is in doing this for a platform where weight and power are strictly limited and where retrieval is problematic. We are proceeding in an incremental manner, where each step serves as a guide for later development. Our first payload was a simple Styrofoam box housing a Canon Ixus 115HS camera. Styrofoam was used for the payload box primarily because it was lightweight, easy to handle and to cut, and readily available. It also provides a good insulation for high-altitude flights. A photograph of the box attached to the balloon is shown in Figure 1 (left). We began by using a 9 mm thick nylon rope, but switched to a lightweight plasticized nylon rope with mass of 200 g/m and 3 mm thickness. This camera was chosen due to its compact size, light weight (~ 140 g, including batteries and memory card) and 12.1 megapixel CMOS sensor with full-HD quality video-recording capability. In future flights we will use Go Pro camera because it has a wider field-of-view (170°) and is suitable for experiments where vibrations and movements are involved. The second payload had the same camera and the miniaturized universal data logger (MSR145). The third payload had in addition the GPS unit.

The data logger consists of the pressure sensor, temperature sensor, humidity sensor with integrated temperature sensor and three-axis accelerometer ($X$, $Y$ and $Z$ axes) for continuous sensing and storage of atmospheric parameters such as pressure, temperature and relative humidity (RH), together with three-axis acceleration. The data logger was programed to record the readings every 1 sec. The compactness ($20 \times 15 \times 52$ mm), light weight (~ 16 g) and durable design of the data logger makes it ideally suited for routine monitoring of the environment the balloon passes through[11]. The data logger is equipped with software to plot and calculate the atmospheric data. We have calibrated the temperature sensor of the data logger using the following steps:

1. A standard laboratory oven was used, which could be heated to specified temperatures, with maximum temperature of 200°C, heating resistance 37 ohm, heater voltage 45 V.
2. Our temperature sensor, along with high accuracy temperature sensor (DS1621) and a digital thermometer, was placed in the oven and heated to different temperatures.
3. The values for different temperatures were noted, compared and it was found that the data logger temperature sensor performed satisfactorily.

The GPS unit of Satguide tracker brand was chosen due to its small size ($64 \times 46 \times 17$ mm) and light weight (50 g). It is a GSM-based unit which sends information back through the mobile network[12]. The GPS was programmed to give location of the balloon every 10 sec.

The most expensive part of the payload is currently the data logger (Table 1). We are trying to cut down on the total expenditure involving the payload components by experimenting with building the small flight computer using BASIC-Stamp-2 microcontroller (BS2), which will be interfaced with a cheaper GPS receiver VPN1513, altimeter MS5607 to measure the atmospheric parameters (instead of the data logger), and the gyroscope module three-axis L3G4200D to stabilize the platform. The data will be transferred to a USB device (for example, a pen-drive) via a memory stick data logger from BS2. The cost of the flight computer is Rs 26,179 and we estimate the cost of our first free-floating flight (1.2 kg balloon) at about Rs 50,000, with Canon Ixus 115HS camera on-board the payload.

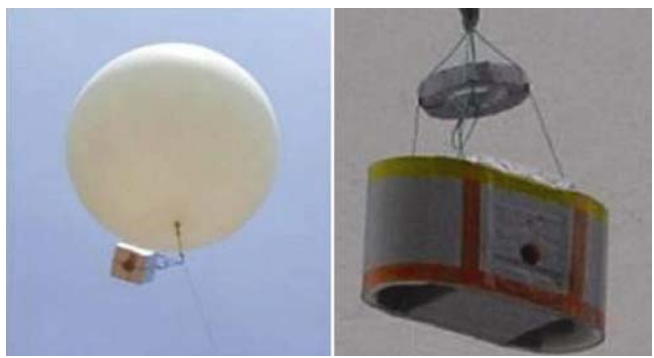

**Figure 1.** First and second flight payload models.

Table 1. Total expenditure to date

| Equipment | Cost (Rs) | | |
|---|---|---|---|
| Hardware | | | |
|     Cylinder (2) | 34,000 | | |
|     Gas regulator (2) | 17,800 | | |
| Payload components | | | |
|     Camera (3) | 49,950 | | |
|     Data logger | 88,027 | | |
|     GPS unit | 10,106 | | |
| Accessories | | | |
|     (Styrofoam, wires, rope, etc.) | 9,124 | | |
| Total | 209,007 | | |
| Expendables | First flight | Second flight | Third flight |
| Helium | 3,944 | 3,944 | 8,120 |
| 1.2 kg balloon | 2,345 | 2,345 | – |
| 3 kg balloon | – | – | 18,520 |
| Total | 6,289 | 6,289 | 26,640 |

## Tethered flights

A set of approvals from various government agencies and nearby airports is necessary for a free-flying balloon in India. We expect to receive these approvals shortly and have used the intervening time for experimenting with tethered flights. The tethered flights consisted of three flights. The first flight aimed at getting acquainted with the pre-launch and launch activities using a 1.2 kg balloon (Figure 1, left). The second flight aimed at improving the structure of the payload box, studying the profile of low troposphere and testing the data-recording capability of the data logger. The third flight was carried out with a 3 kg balloon to achieve greater altitudes compared to the 1.2 kg balloon. This allowed us to go up to heights of about 400 m with flight durations of several hours and test the conditions we expect in a free-floating balloon. We have found that the most difficult part of the experiment design was to ensure the stability of the payload box, which is subject to winds and other random accelerations. We are now using a rectangular box with rounded edges made out of Styrofoam and plastic fibre (Figure 1, right). Such a structure ensures alignment in one particular direction of the wind, so that rotation of the payload is minimized.

### First flight (5 March 2012)

On our first tethered flight the payload box contained the Canon Ixus 115HS digital camera (Figure 1, left). The total weight of the payload was 300 g, and we used 1.2 kg balloon, which we filled with helium to about 70% capacity. The conclusions from our debut flight are as follows:

1. The rope used for tethering was thick and heavy, due to which the balloon's altitude was only about 60–70 m (judging by the rope length).

2. The cubic-shaped Styrofoam box had resulted in too many drifts and rotations (which we could see from the videos recorded by the camera).

### Second flight (11 June 2012)

The second experiment was to overcome the drawbacks of the first flight. For the second flight, we changed the shape of the payload box to rectangular with rounded edges to align the payload in one particular direction, along the wind, so that the sideways movement of the payload is limited. The box was open at the top and bottom, allowing movement of air through it (Figure 1, right). This payload was much smaller and lighter than the first one. The nylon rope used for this flight was 3 mm thick and lightweight. The conclusions from this flight are as follows:

1. The new rope ensured a height of about 150 m. The average altitude of CREST campus is 940 m and the maximum height attained by the balloon in the second flight was ~145 m. The altitude was calculated from the pressure data as follows:

$$z = \frac{T_0}{\tau}\left[1 - \left(\frac{P_{\text{local}}}{P_{\text{std}}}\right)^{R\tau/g}\right], \quad (1)$$

where $P_{\text{local}}$ is the measured pressure by the data logger in mbar, $P_{\text{std}} = 1013.25$ mbar is the standard sea-level pressure, $R = 287.04$ J/kg/K is the gas constant for dry air, $g = 9.801$ m/s$^2$ is the acceleration due to gravity, $\tau = 0.0065$ K/m is the lapse rate of temperature with altitude and $T_0 = 288$ K is the ground temperature.

2. The data logger has given satisfactory results, and recorded temperature, pressure and three-axis acceleration throughout the flight. Figure 2 (top) shows the profile of temperature and RH versus time. It has been observed that temperature and RH have variations during the pre- and post-flight period (duration 0–500 sec and 4400–4600 sec) and almost constant readings (30–35°C for temperature and 36–38% for RH) during the flight (600–3516 sec). RH is defined as the ratio of vapour pressure to the saturation vapour pressure. Air is saturated when RH is 100%. RH has an inverse relationship with temperature; such that when temperature is the highest, RH is the lowest and when temperature is at a minimum, RH is highest[13]. This is assuming that actual vapour content of the air is not changing. This behaviour is seen in the temperature and humidity profiles in Figure 2 (top), where temperature curve (red) and humidity curve (blue) follow a near-exact mirror profile. The acceleration profiles from the three-axis accelerometer are shown in Figure 2 (bottom). The X, Y and Z-axes acceleration data are plotted on the same graph with three different colours. The accelerometer is calibrated in such a way that the Z-axis acceleration is directed vertically, thus it is shifted by 1 *g* up the scale relative to *x* and *y* values. The acceleration profile shows up spikes up to a maximum of 9 *g* amplitude. We attribute this to high-velocity winds noticed on the day the experiment was carried out. This behaviour is the same as in the third flight (Figure 3, bottom).

*Third flight (12 July 2012)*

A 3 kg balloon was used for this experiment. The payload in the third flight consisted of a Canon Ixus115HS camera, MSR145 data logger and the GPS unit. The shape of the payload box was retained as in the second flight with the closed structure being the only difference. This flight was longer with the duration of nearly 3 h. The following conclusions were drawn:

1. As expected with the new rope, a balloon altitude of 350–400 m was successfully achieved. The altitude was calculated using eq. (1) from pressure data and the variation of altitude with time of the flight is shown in Figure 4 (top).
2. The pressure versus time profile (Figure 3, top) follows an expected trend with large fluctuations in the mid-flight, which we attribute to the atmospheric turbulence at altitudes of 200–300 m (on that day the winds were very strong). The same behaviour was observed in the second flight, where fluctuations were smaller as the height achieved was lower.
3. The acceleration profiles from the three-axis accelerometer are shown in Figure 3 (bottom). The *X*, *Y* and

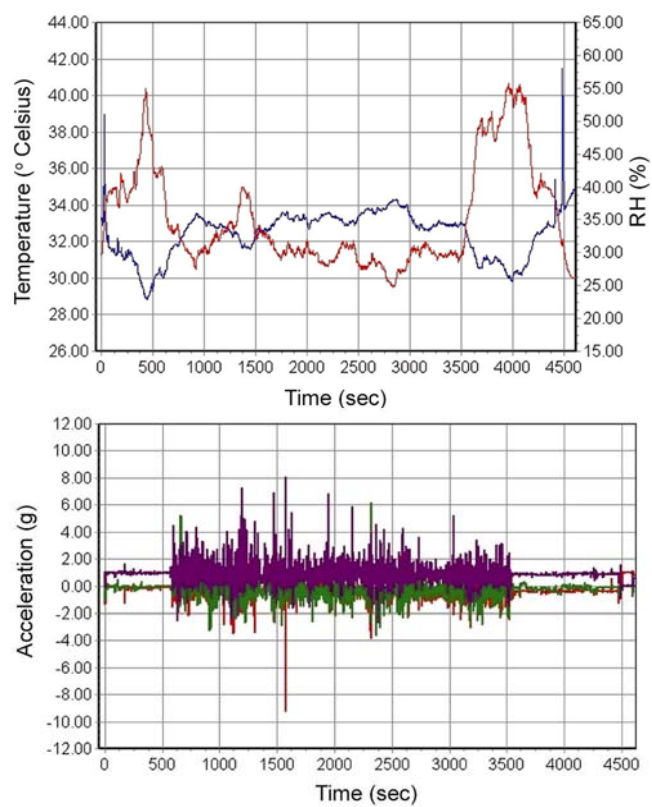

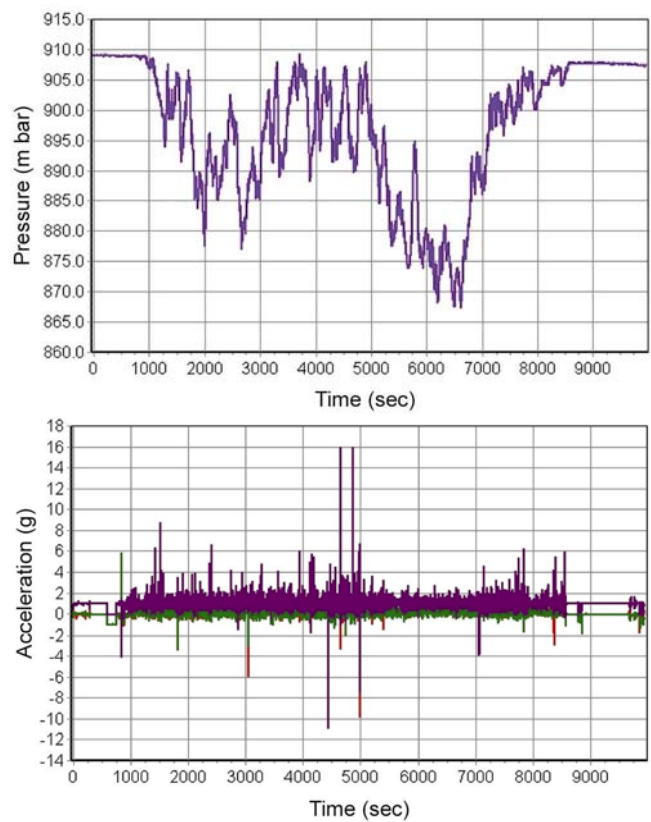

**Figure 2.** (Top) Variation of temperature and relative humidity with time. *Y*-axis on the left is temperature (°C), *Y*-axis on the right is humidity (%). Launch time (release of the balloon) was at 600 sec and the flight was terminated at 3516 sec (payload was brought to the ground). The temperature curve (red) and relative humidity curve (blue) follow the near exact mirror profile. The large fluctuations in humidity level at the starting and ending time of the flight may be attributed to human handling of the humidity sensor. (Bottom) Variation of three-axis acceleration with time. *Y*-axis is acceleration (*g*). Red curve corresponds to *X*-axis acceleration, green to *Y*-axis and purple to *Z*-axis accelerations. Launch and termination times as in the top panel. The zero of the *Z*-axis is shifted with respect to *X* and *Y* by 1 *g*. During the flight the rms for all the accelerations is about 0.4204 *g*, while rms at rest (at the beginning and at the end of the experiment) is about 0.04 *g*. The large spikes during the flight are most probably due to random disturbances of the payload (strong winds).

**Figure 3.** (Top) Pressure variation with time. *Y*-axis is pressure (mbar). Launch time was at 800 sec and the flight was terminated at 8400 sec. The pressure at the starting and ending time of the flight is consistent with the standard atmospheric pressure at the CREST campus location at the ground level. Large fluctuations during the flight may be attributed to the atmospheric turbulence at heights of 200–400 m with the minimum pressure (867 mbar) reached at the highest altitude of 400 m. (Bottom) Variation of three-axis acceleration with time. *Y*-axis is acceleration (*g*). Red is *X*-axis acceleration, green is *Y*-axis and purple is *Z*-axis acceleration. The large spikes during the flight can be attributed to random disturbances of the payload (strong winds).

Z-axes acceleration data are plotted on the same graph with three different colours. The acceleration profile shows spikes up to a maximum amplitude of 15 g in the Z direction, which is to be expected, as it is in the vertical direction that the action of forces of gravity, buoyancy and drag is most prominent. We attribute these spikes to high-speed winds noticed on the day the experiment was carried out. The accelerometer was calibrated in the lab where at rest the total rms of fluctuations was about 0.04 in all three axes.

4. The temperature profile as a function of altitude is shown in Figure 4 (top, red dots). There is an initial increase in temperature due to the payload being exposed to the midday sun followed by a decrease in the temperature with altitude. The rate at which the temperature decreases, $dt/dz$, is called the environmental lapse rate. On the same plot, in black dots, we also show the dependence of pressure on the altitude. Atmospheric pressure is maximum at sea level and decreases with altitude. This is shown by the pressure–altitude variation in both the second and third flights.

5. The GPS device was programmed to send its location at an interval of every 10 sec to a mobile number. Figure 5 shows the path followed by the balloon from the GPS data plotted on Google Map using a KML program[14]. The code and steps to program the device are available on the IIA Balloon Group website at http//www.iiap.res.in/balloon/.

**Future plans**

We plan to fly a spectrograph on a free-floating balloon, which will enable observations in near-UV at altitudes of 30–40 km. The spectrograph will be a compact-design, three-bounce system with mirrors and toroidal grating[15]. The stability of the payload box is an essential factor and we plan to use gyroscopes to stabilize the payload in all three directions. We intend to have a parachute attached to the payload box to ensure safe descent to the ground. The detachment of the parachute will be done by a flight termination unit. The flight computer will serve as an intelligent system on-board, collecting data on atmospheric parameters such as temperature, pressure and humidity and providing power to the payload. We are also contemplating performing high-altitude astrobiology experiments, such as for example survival of microbes in the upper atmosphere[16]. Computer simulations for the ascent of free-floating balloon and the descent of the payload and predictions of the flight path, as well as atmospheric parameters during the flight, are being conducted, and the results will be presented in a forthcoming paper at the ASI conference at Thiruvananthapuram in 2013.

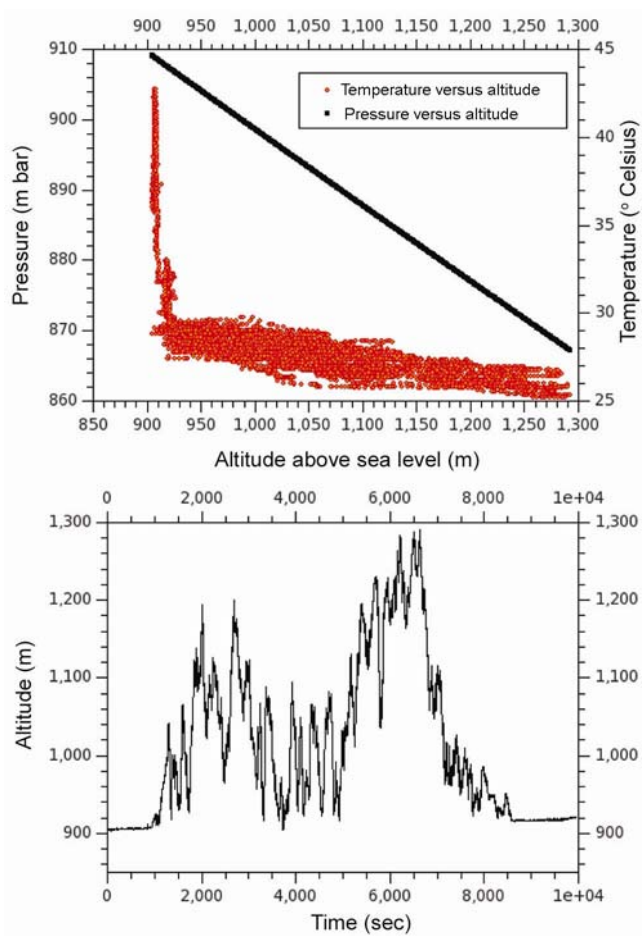

**Figure 4.** (Top) The third flight profiles of temperature (red dots) and pressure (black dots) versus altitude. Left Y-axis is pressure (mbar), right Y-axis is temperature (°C). The temperature in the troposphere decreases with altitude, which can be seen in the figure. The pressure of the atmosphere decreases with altitude. It is maximum at sea level; at the altitude of CREST campus it was ~ 900 mbar on the date of flight. Altitude values were calculated from the pressure data using eq. (1). (Bottom) Altitude variation with time. Altitude was calculated from the pressure data using eq. (1).

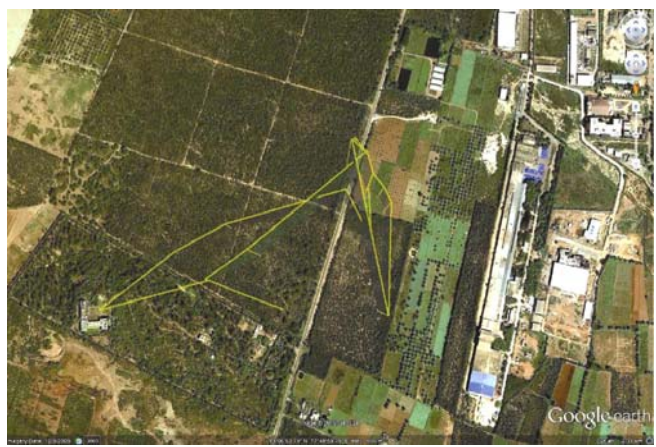

**Figure 5.** Trajectory followed by the balloon obtained from the GPS data at the location of the third experiment.

**Summary**

We have presented the description and results of the three tethered experiments carried out in the Centre for Research and Education in Science and Technology (CREST) campus, IIA, Hosakote. The data logger used in these experiments gave satisfactory results and we could study the atmospheric profile at altitudes from about 100 to 400 m. The GPS locations were provided by the tracker every 10 sec and helped us to follow the balloon path. The camera helped in capturing the video for 1 h duration. The atmospheric interdependent parameter (pressure, temperature, altitude) behaviour was studied in the lower troposphere using tethered flights. The repeatability of the tethered flight results needs to be verified. We are anticipating the set of approvals for free-floating balloons soon to begin stratospheric observations.

ACKNOWLEDGEMENTS. We thank Hindustan Aeronautics Ltd, Chennai Airport Authorities, and Jakkur Aerodrome for providing the necessary clearances to carry out the free-floating balloon experiments. We also thank Srinivasan (CATCO Officer-Air Force Station Training Command Headquarters, Bangalore) for his help in obtaining these clearances.

Received 18 December 2012; accepted 28 January 2013